\documentclass[runningheads]{llncs}
\usepackage{graphicx}
\usepackage{amsmath}
\usepackage{amssymb}
\usepackage{hyperref}
\usepackage{color}

\usepackage{algorithm}
\usepackage{algpseudocode}
\usepackage{booktabs}
\usepackage{todonotes}

\newcommand{\expSym}{\mathbb{E}}
\newcommand{\E}[1]{\ensuremath{\expSym\left(#1\right)}}

\newcommand{\T}{\top}

\newcommand{\Cov}[2]{\ensuremath{\mathrm{Cov}(#1, #2)}}
\newcommand{\vw}[1]{}

\renewcommand{\inst}[1]{\textsuperscript{#1}}

\DeclareMathOperator*{\argmin}{arg\,min}

\newtheorem{model}{Model}

\begin{document}
\title{Control Variates for Stochastic Simulation\\
of Chemical Reaction Networks}
\author{Michael Backenk\"ohler\inst{1}, Luca Bortolussi\inst{1,2}, Verena Wolf\inst{1}}
\authorrunning{M.\ Backenk\"ohler et al.}
\institute{\inst{1}Saarland University, Germany,\ \ \inst{2}University of Trieste, Italy}
\maketitle
\begin{abstract}
Stochastic simulation is a widely used method for 
estimating quantities in models of chemical 
reaction networks where uncertainty plays a crucial role.
However,  reducing  the statistical uncertainty of the corresponding  
estimators requires the generation of a large number 
of simulation runs, which is computationally expensive.
To reduce the number of necessary runs, we propose a
variance reduction technique based on control variates.
We exploit  constraints on the statistical moments of the 
stochastic process
to reduce the estimators' variances.
We develop an algorithm that selects appropriate control
variates in an on-line fashion and demonstrate the efficiency of our approach  on several case studies.

\end{abstract}
\section{Introduction}
Chemical reaction networks that are used to describe cellular processes are often subject to inherent stochasticity.
The dynamics of gene expression, for instance, is influenced by 
single random events (e.g.\ transcription factor binding) and 
hence, models that take this randomness into account must monitor
discrete molecular counts and reaction events that change these counts.
Discrete-state continuous-time Markov chains have successfully  been
used to describe  networks of chemical reactions
over time that correspond to the basic events of such processes. 
The time-evolution of the corresponding probability distribution is 
given by the chemical master equation, whose numerical solution is
extremely challenging because of the enormous size of the underlying
state-space. 

Analysis approaches based on sampling, such as the Stochastic Simulation Algorithm (SSA) \cite{gillespie77}, can  be applied
independent of the size of the model's state-space. 
However, statistical approaches are costly since a large number
of simulation runs is necessary to reduce the statistical 
inaccuracy of estimators. This problem is particularly severe
if reactions occur on multiple time scales or if the event of interest is rare.
A particularly popular technique to speed up simulations is $\tau$-leaping which applies
multiple reactions in one step of the simulation.
However, such multi-step simulations rely on certain assumptions about
the number of reactions in a certain time interval. These assumptions
are typically only approximately fulfilled and therefore introduce  approximation
errors on top of the statistical uncertainty of the considered point estimators.

Moment-based techniques offer a fast approximation of the statistical moments of the model. 
The exact moment dynamics can be expressed as an infinite-dimensional system of ODEs, which cannot be directly  integrated for a transient analysis.
Hence, ad-hoc approximations need to be introduced, expressing higher order moments as functions of lower-order ones \cite{ale2013general,engblom2006computing}.
However, moment-based approaches rely on assumptions about the dynamics that are
often not even approximately fulfilled and may lead to high approximation errors. 
Recently, equations expressing the moment dynamics have also been used as constraints for parameter
estimation \cite{backenkohler2018moment} and for computing moment bounds using semi-definite programming \cite{dowdy2018dynamic,ghusinga2017exact}.

In this work, we propose a combination of such moment constraints with the SSA approach.
Specifically, we   interpret these constraints as random variables 
that are correlated with the estimators of interest usually given as functions of chemical population variables.
These constraints can be used  as  (linear) control variates in order to improve the final estimate and reduce its variance \cite{lavenberg1982statistical,szechtman2003control}.
The method is easy on an intuitive level: If a control variate
is positively correlated with the function to be estimated then 
we can use the estimate of the variate to
adjust the target estimate.

The incorporation of control variates into the SSA
introduces   additional simulation costs for the calculation of the constraint values.
These values are integrals over time, which we accumulate  based on the
piece-wise constant trajectories.
This introduces a trade-off between the variance reduction that is achieved
by using control variates versus the increased simulation cost.
This trade-off is expressed as the product of the variance reduction ratio
and the cost increase ratio.

For a good trade-off, it is crucial to find an appropriate set of control variates.
Here we propose a class of constraints which is parameterized by a moment vector
and a weighting parameter, resulting in infinitely many choices.
We present an algorithm that samples from the set of all constraints and
proceeds to remove constraints that are either only weakly correlated with the
target function or are redundant in combination with other constraints.

In a case study, we explore different variants of this algorithm
both in terms of generating the initial constraint set and of
removing  weak or redundant constraints.
We find that the algorithm's efficiency is superior to a standard estimation procedure
using stochastic simulation alone
in almost all cases.

Although in this work we focus on  estimating first order moments at
fixed time points, the proposed approach can in principle deal with any property that can
be expressed in terms of expected values such as probabilities
of complex path properties.
Another advantage of our technique is that an increased efficiency is achieved without the price of an additional approximation error as it is the case for methods based on moment approximations or multi-step simulations.

This paper is structured as follows. In Section~\ref{sec:related} we give a brief
survey of  methods and tools related to efficient stochastic simulation
and moment techniques. In Section~\ref{sec:bg} we introduce the
common stochastic semantics of chemical reaction networks. From these
semantics we show in Section~\ref{sec:moments} how to derive
constraints on the moments of the transient distribution.
The variance reduction technique of control variates is described in
Section~\ref{sec:var_red}.
We show the design of an algorithm using moment constraints to reduce
sample variance in Section~\ref{sec:algo}.
The efficiency and other characteristics of this algorithm are evaluated
on four non-trivial case studies in Section~\ref{sec:study}.
Finally, we discuss the findings and give possibilities for further
work in Section~\ref{sec:conclusion}.

\section{Related Work}\label{sec:related}
Much research has been directed at the efficient analysis of stochastic chemical reaction
networks.
Usually research focuses on improving efficiency by making certain approximations.

If the state-space is finite and small enough one can deal with the underlying Markov
chain directly.
But there are also  cases where the transient distribution has an infinitely large support
and one can still deal with explicit state probabilities.
To this end, one can fix a finite state-space, that should contain most of the
probability~\cite{munsky2006finite}. Refinements of the method work
dynamically and adjust the state-space according to the transient
distributions~\cite{andreychenko2011parameter,henzinger2009sliding,mateescu2010fast}.

On the other end of the spectrum there are mean-field approximations, 
which model the mean densities faithfully in the system size limit~\cite{bortolussi2013continuous}.
In between there are techniques such as moment closure \cite{singh2006lognormal}, that
not only consider the mean, but also the variance and other higher order moments.
These methods depend on ad-hoc approximations of higher order moments to
close the ODE system given by the moment equations.
Yet another class of methods approximate molecular counts continuously and approximate the dynamics in such a continuous space, e.g. the system size expansion~\cite{van1992stochastic} and the
chemical Langevin equation~\cite{gillespie2000chemical}.

While the moment closure method uses ad-hoc approximations for high order moments to
facilitate numerical integration, they can be avoided in some contexts.
For the equilibrium distribution, for example, the time-derivative of all moments is equal to zero.
This directly yields constraints that have been used for parameter estimation at
steady-state~\cite{backenkohler2018moment}
and bounding moments of the equilibrium distribution using semi-definite
programming~\cite{ghusinga2017estimating,ghusinga2017exact,kuntz2017rigorous}.
The latter technique of bounding moments has been successfully adapted in the 
context of transient analysis~\cite{dowdy2018dynamic,sakurai2017convex,sakurai2019bounding}.
We adapt the constraints proposed in these works to improve statistical estimations via stochastic
simulation (cf.\ section~\ref{sec:moments}).

While the above techniques give a deterministic output, stochastic simulation generates
single executions of the stochastic process~\cite{gillespie77}.
This necessitates accumulating large numbers of simulation runs to estimate
quantities.
This adds a significant computational burden. Consequently, some effort
has been directed at lowering this cost.
A prominent technique is $\tau$-leaping~\cite{gillespie2001approximate},
which in one step performs multiple instead of only a single reaction.
Another approach is to find approximations that are specific to the problem at hand,
such as approximations based on time-scale separations~\cite{cao2005slow,bortolussi2015efficient}.

Recently, multilevel Monte Carlo methods have been applied in to time-inhomogenous
CRNs~\cite{anderson2018low}. In this techniques estimates are combined
using estimates of different approximation levels.


The most prominent application of a variance reduction technique in the context
of stochastic reaction networks is importance sampling~\cite{kuwahara2008efficient}.
This technique relies on an alteration of the process and then weighting samples
using the likelihood-ratio between the original and the altered process.

\section{Stochastic Chemical Kinetics}\label{sec:bg}
A chemical reaction network (CRN) describes the interactions between
a set of species $S_1,\dots, S_{n_S}$ in a well-stirred reactor.
Since we assume that all reactant molecules are spatially uniformly distributed,
we just keep track of the overall amount of each molecule.
Therefore the state-space is given by $\mathcal{S}\subseteq\mathbb{N}^{n_S}$.
These interactions are expressed a set of \emph{reactions} with a certain
inputs and outputs, given by the vectors $v_j^{-}$ and $v_j^{+}$ for
reaction $j=1,\dots,n_R$, respectively. Such reactions are denoted as
\begin{equation}\label{eq:reaction}
    \sum_{i=1}^{n_S} v_{ji}^{-} S_i
    \xrightarrow{c_j}
    \sum_{i=1}^{n_S} v_{ji}^{+} S_i\,.
\end{equation}
The reaction rate constant $c_j>0$ gives us information on the propensity
of the reaction.
If just a constant is given, \emph{mass-action} propensities are assumed.
In a stochastic setting for some state $x\in\mathcal{S}$ these are
\begin{equation}\label{eq:stoch_mass_action}
    \alpha_j(x)=c_j\prod_{i=1}^{n_S}\binom{x_i}{v_{ji}^{-}}\,.
\end{equation}
The system's behavior is described by a stochastic
process $\{X_t\}_{t\geq 0}$.
The propensity function gives the infinitesimal probability
of a reaction occurring, given a state $x$. That is, for a
small time step $\delta t >0$
\begin{equation}\label{eq:reaction_prob}
    \Pr(X_{t+\delta t}=x+v_j\mid X_t=x)
    =\alpha_j(x)\delta t + o(\delta t)\,.
\end{equation}
This induces a corresponding continuous-time
Markov chain (CTMC) on $\mathcal{S}$ with generator matrix\footnote{Assuming a fixed enumeration of the state space.}
\begin{equation}\label{eq:cme_generator}
    Q_{x,y} = \begin{cases}
        \sum_{j:x+v_j = y}\alpha_j(x)\,,&\text{if}\;x\neq y\\
        -\sum_{j=1}^{n_R} \alpha_j(x)\,, &\text{otherwise.}
    \end{cases}
\end{equation}
Accordingly, the time-evolution of the process' distribution,
given an initial distribution $\pi_0$, is given by the
Kolmogorov forward equation, i.e.\ $\frac{d\pi_t}{dt}=Q\pi_t$, where $\pi_t(x)=\Pr(X_t=x)$.
For a single state, it is commonly referred to 
as the \emph{chemical master equation} (CME)
\begin{equation}\label{eq:cme}
    \frac{d}{d t} \pi_t(x) =
    \sum_{j=1}^{n_R}\left(
        \alpha_j(x-v_j)\pi_t(x-v_j) - \alpha_j(x)\pi_t(x)
    \right)\,.
\end{equation}
A direct solution of \eqref{eq:cme} is usually not possible.
If the state-space with non-negligible probability is suitably small, a state space
truncation could be performed. That is, \eqref{eq:cme} is integrated on a possibly time-dependent subset
$\hat{\mathcal{S}}_t\subseteq\mathcal{S}$ \cite{henzinger2009sliding,munsky2006finite,spieler2014numerical}.
Instead of directly analyzing \eqref{eq:cme}, one often resorts to simulating trajectories.
A trajectory $\tau=x_0t_1x_1t_1\dots t_n x_n$ over the interval $[0,T]$ is a sequence of states $x_i$
and corresponding
jump times $t_i$, $i=1,\dots,n$ and $t_n=T$.
We can sample trajectories of $X$ by using stochastic simulation~\cite{gillespie77}.

Consider the birth-death model below as an example.
\begin{model}[Birth-death process]\label{model:bd}
    A single species $\mathsf{A}$ has a constant
    production and a decay that is linear in the current amount of molecules.
    Therefore the model consists of two mass-action reactions
    $$\varnothing \xrightarrow{\gamma} \mathsf{A}\,,\quad
    \mathsf{A}\xrightarrow{\delta}\varnothing\,,$$
    where $\varnothing$ denotes   no reactant or no product,
    respectively.
\end{model}
For Model~\ref{model:bd} the change of probability mass in a single state $x>0$ is described by expanding
\eqref{eq:cme} and
$$\frac{d}{dt}\pi_t(x)=\gamma \pi_t(x-1) + \delta \pi_t(x+1) - (\gamma + \delta)\pi_t(x)\,.$$
We can generate trajectories of this model by choosing either reaction, with a probability that is
proportional to its rate given the current state $x_i$.
The jump time $t_i- t_{i+1}$ is determined by sampling from an exponential distribution with rate $\gamma+x_i\delta$.


\section{Moment Constraints}\label{sec:moments}
The time-evolution of $\E{f(X_t)}$ for some function $f$ can be directly derived from \eqref{eq:cme} by 
computing the sum
$$\sum_{x\in\mathcal{S}}f(x)\frac{d}{dt}\pi_t(x)\,,$$ which yields 
\begin{equation}\label{eq:mom_ode}
    \frac{d}{dt}\E{f(X_t)} = \sum_{j=1}^{n_R}\E{\left(f({X_t + v_j}) - f(X_t)\right)\alpha_j(X_t)}\,.
\end{equation}
While many choices of $f$ are possible, for this work we will restrict ourselves to
monomial functions $f(x)=x^m$, $m\in\mathbb{N}^{n_S}$
i.e.\ the \emph{non-central moments} of the process.
The \emph{order} $\lvert m\rvert$ of a moment $\E{X^m}$ is the sum over the exponents,
i.e.\ $\lvert m\rvert =\sum_im_i$.
The integration of \eqref{eq:mom_ode} with such functions $f$ is well-known in the context of
moment approximations of CRN models.
For most models the arising ODE system is infinitely large, because the time-derivative of
low order moments usually depends on the values of higher order moments.
To close this system, \emph{moment closures}, i.e.\ ad-hoc approximations of higher order moments
are applied \cite{schnoerr2015comparison}.
The main drawback of this kind of analysis is that it is not known whether the chosen closure gives
an accurate approximation for the case at hand.
Here, such approximations are not necessary, since we will apply the moment dynamics in the context
of stochastic sampling instead of trying to integrate \eqref{eq:mom_ode}.

Apart from integration strategies,
setting \eqref{eq:mom_ode} to zero has been used as a constraint for parameter estimation at steady-state
\cite{backenkohler2018moment} and bounding moments at steady-state~\cite{dowdy2018bounds,ghusinga2017exact,kuntz2017rigorous}.
The extension of the latter has recently lead to the adaption of these constraints
to a transient setting~\cite{dowdy2018dynamic,sakurai2019bounding}.
These two transient constraint variants are analogously derived by multiplying \eqref{eq:mom_ode}
by a time-dependent, differentiable weighting function $w(t)$ and integrating:

Multiplying with $w(t)$ and integrating on $[t_0, T]$ yields~\cite{dowdy2018dynamic,sakurai2019bounding}
\begin{equation}\label{eq:poly_con}
\begin{split}
        & w(T)\E{f(X_{T})}
        - w(t_0)\E{f(X_{t_0})}
        - \int_{t_0}^{T}\frac{dw(t)}{dt}\E{f(X_t)}\,dt\\
        =&\sum_{j=1}^{n_R}\int_{t_0}^{T}w(t)
        \E{\left(f{(X_t + v_j)} - f(X_t)\right)\alpha_j(X_t)}\,dt
        \end{split}
\end{equation}

In the context of computing moment bounds via semi-definite programming
the choices $w(t)=t^s$~\cite{sakurai2019bounding} and $w(t)=e^{\lambda(T - t)}$~\cite{dowdy2018dynamic}
have been proposed.
While both choices proved to be effective in different case studies, relying solely on the latter choice,
i.e.\ $w(t)=e^{\lambda(T - t)}$ was sufficient.

By expanding the rate functions and $f$ in \eqref{eq:poly_con} and substituting the
exponential weight function we can re-write \eqref{eq:poly_con} as
\begin{equation}\label{eq:simpl_exp}
        0 =\,
         \E{f(X_{T})}
        - e^{\lambda T}\E{f(X_{t_0})}
        + \sum_{k}c_k\int_{t_0}^{T}e^{\lambda(T-t)}\E{X_t^{m_k}}\,dt
\end{equation}
with coefficients $c_k$ and vectors $m_k$ defined accordingly.
Assuming the moments remain finite on $[0,T]$, we can define the random variable
\begin{equation}\label{eq:z}
        Z =\,
         f(X_{T})
        - e^{\lambda T}f(X_{t_0})
        + \sum_{k}c_k\int_{t_0}^{T}e^{\lambda(T-t)}X_t^{m_k}\,dt
\end{equation}
with $\E{Z}=0$.

Note, that a realization of $Z$ depends on the whole trajectory $\tau=x_0t_1 x_1 t_1 \dots\allowbreak t_n x_n$ over $[t_0,T]$.
Thus, for the integral terms
in \eqref{eq:z} we have to compute sums
\begin{equation}\label{eq:dis_int}
    \frac{1}{\lambda}\sum_{i=1}^n\left(e^{\lambda(T - t_{i+1})}
    - e^{\lambda(T-t_i)}\right)x_i^{m_k}\,,
\end{equation}
over a given trajectory.
This accumulation is best done during the simulation to avoid storing the whole trajectory.
Still, the cost of a simulation run increases.
For the method to be efficient, the variance reduction (Section~\ref{sec:var_red}) needs
to overcompensate for this increased cost of a simulation run.

For Model~\ref{model:bd} the moment equation for $f(x)=x$ becomes
$$\frac{d}{dt}\E{X_t}=\gamma - \delta\E{X_t}\,.$$
The corresponding constraint \eqref{eq:simpl_exp} with $\lambda=0$ gives
$$0=\E{X_T} - \E{X_0} - \gamma T + \delta \int_0^{T} \E{X_t}\,dt\,.$$
In this instance the constraint  leads to an explicit function of
the moment over time. If  $X_0=0$ w.p.\ 1, then \eqref{eq:simpl_exp} becomes
\begin{equation}\label{eq:bd_constraint}
\E{X_T} = \frac{\gamma}{\delta} \left(1 - e^{-\delta T}\right)
\end{equation}
when choosing $\lambda=-\delta$.

\section{Control Variates}\label{sec:var_red}
Now, we are interested in the estimation of some quantity $\E{V}$
by stochastic simulation.
Let $V_1,\dots,V_n$ be independent samples of $V$.
Then the sample mean $\hat{V}_n
=\frac{1}{n}\sum_{i=1}^n V_k$ is an estimate of $\E{V}$.
By the central limit theorem
\[
\sqrt{n}\hat{V}_n\xrightarrow{d}N(\E{V},\sigma_V^2)\,.
\]
Now suppose, we know of a random variable $Z$ with $0=\E{Z}$.
The variable $Z$ is called a \emph{control variate}.
If a control variate $Z$ is correlated with $V$, we can
use it to
reduce the variance of $\hat{V}_n$~\cite{glasserman2005large,nelson1990control,szechtman2003control,wilson1984variance}.
For example, consider we are running a set of simulations and consider a single
constraint.
If the estimated value of this constraint is larger than zero and we estimate a positive correlation
between the constraint $Z$ and $V$, we would, intuitively, like to {decrease} our
estimate $\hat{V}_n$ accordingly.
This results in an estimation of the mean of the random variable $$Y_{\beta}= V-\beta Z$$ instead of $V$.
The variance
$$\sigma_{Y_{\beta}}^2 = \sigma_V^2-2\beta \Cov{V}{Z} + \beta^2\sigma_Z^2\,.$$
The optimal choice $\beta$ can be computed by  considering the minimum of $\sigma_{Y_\beta}^2$. Then
$$\beta^{*}={\Cov{V}{Z}}/{\sigma_Z^2}\,.$$
Therefore $\sigma_{Y_{\beta^{*}}}=\sigma_Z^2(1 - \rho_{VZ}^2)$,
where $\rho_{VZ}$ is the correlation of $Z$ and $V$.

If we have multiple control variates, we can proceed in a similar fashion.
Now, let ${Z}$ denote a vector of $d$ control variates and let
\[
\Sigma=
\begin{bmatrix}
\Sigma_{ Z} & \Sigma_{V Z}\\
\Sigma_{ Z V} & \sigma_V^2
\end{bmatrix}
\]
be the covariance matrix of $({Z},V)$.
As above, we estimate the mean of
$
    {Y}_{\beta}=V -{\beta}^{\T}{Z}\,.
$
The ideal choice of $\beta$ is the result of an ordinary least squares regression between $V$
and $Z_i$, $i=1,\dots,n$.
Specifically, $\beta^{*}={\Sigma_{ Z}}^{-1}{\Sigma}_{ Z V}$.
Then, asymptotically
the variance of this estimator is~\cite{szechtman2003control},
\begin{equation}\label{eq:lcv_asym}
    {\sigma_{\hat Y_{\beta^*}}^2} = (1 - R_{ Z V}^2){\sigma_{\hat V}^2}\,, \quad
    R_{ Z V}^2=\Sigma_{ Z V}\Sigma_{ Z}^{-1}\Sigma_{ Z V} / \sigma_V^2\,.
\end{equation}
In practice, however, $\beta^*$ is unknown and needs to be replaced by
an estimate $\hat{\beta}$.
This leads to an increase in the estimator's variance.
Under the assumption of $Z$ and $V$ having a multivariate normal
distribution~\cite{cheng1978analysis,lavenberg1982statistical}, the variance of the estimator is
$\hat{Y}_{\hat{\beta}}=\hat{V}-\hat{\beta}^{\top}\hat{ Z}$
\begin{equation}\label{eq:lcv_norm_varred}
    {\sigma_{\hat{Y}_{\hat{\beta}}}^2} = \frac{n - 2}{n - 2 - d}(1 - R_{ ZV}^2){\sigma_{\hat V}^2}\,.
\end{equation}

Clearly, a control  variate is ``good'' if it is highly correlated with $V$.
The constraint in \eqref{eq:bd_constraint} is an example of the extreme case.
When we use this constraint as a control variate
for the estimation of the mean at some time point $t$, it has a correlation of $\pm1$
since it describes the mean at that time precisely.
Therefore the variance is reduced to zero.
We thus aim to pick control  variates that are highly correlated with $V$.

Consider, for example, the above case of the birth-death process.
If we choose \eqref{eq:bd_constraint} as a constraint, it would always yield
the exact difference of the exact mean to the sample mean and therefore have a 
perfect correlation. Clearly, $\hat\beta$ reduces to 1 and $\hat Y_1 = \E{X_t}$.

\section{Moment-Based Variance Reduction}\label{sec:algo}
We propose an adaptive estimation algorithm (Algorithm~\ref{alg:ssa}) that starts out with
an initial set of control variates
and periodically removes potentially inefficient variates.
The ``accumulator set'' $A$ represents the time-integral terms \eqref{eq:dis_int}.
The size of $A$ has the most significant impact on the overall speed of the algorithm
since it represents the only factor incurring a direct cost increase in the SSA itself (line~\ref{line:run_ssa}).

The algorithm consists of a main loop which performs $n$ simulation runs (line~\ref{line:main_loop}).
Between each run the mean and covariance estimates of $[Z,\;  V]$ are updated (line~\ref{line:updates}).
Every $d<n$ iterations, the control variates are checked for  \emph{efficiency}
and \emph{redundancy} (lines \ref{line:cond_ck_begin}--\ref{line:cond_ck_end}).

\begin{figure}[t]
    \centering
    \hspace{-2.8em}
    \begin{minipage}{.24\textwidth}
    \centering
    \includegraphics[scale=.4]{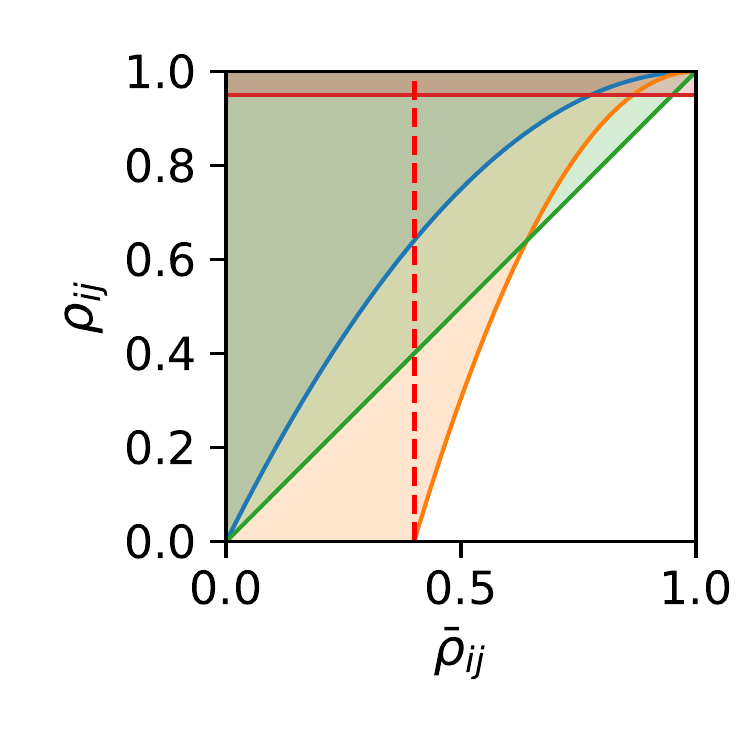}
    \end{minipage}
    \begin{minipage}{.1\textwidth}
    \vfill    \includegraphics[scale=.4]{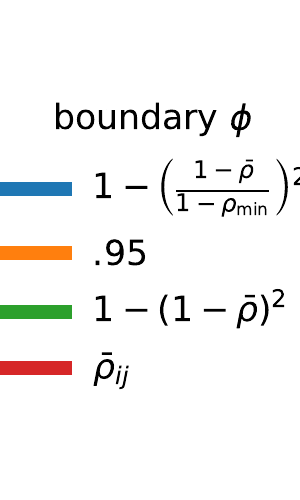} \vfill
    \end{minipage}
    \hspace{1em}
    \begin{minipage}{.29\textwidth}
    \centering
    \includegraphics[scale=0.4]{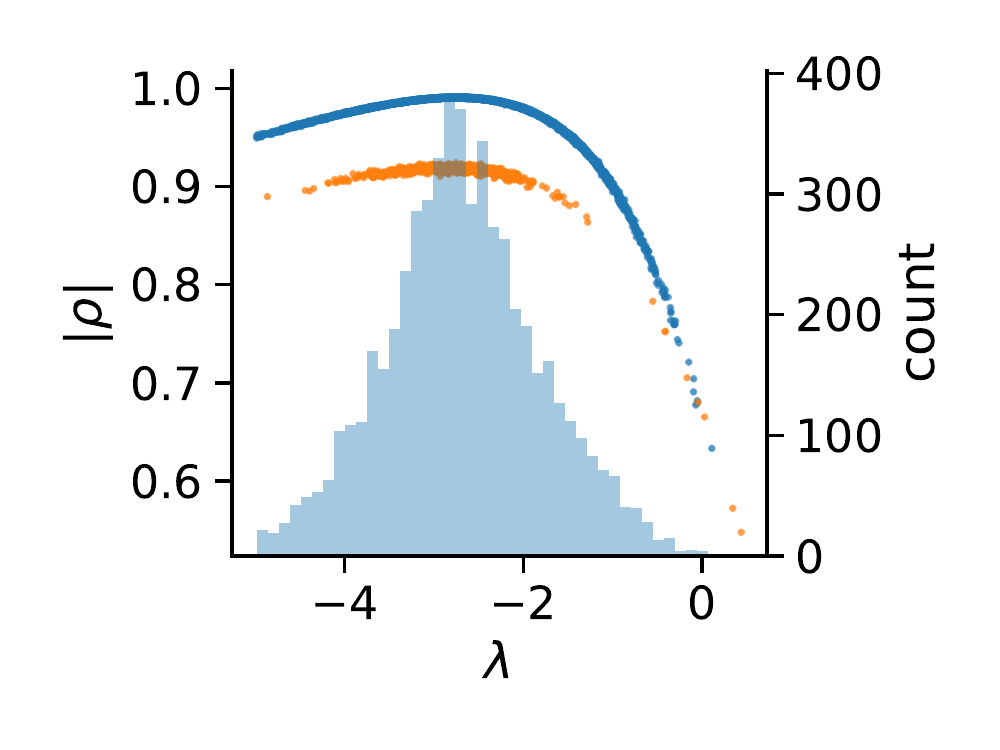}\\
    \end{minipage}
    \hspace{1em}
    \begin{minipage}{.29\textwidth}
    \centering
    \includegraphics[scale=0.4]{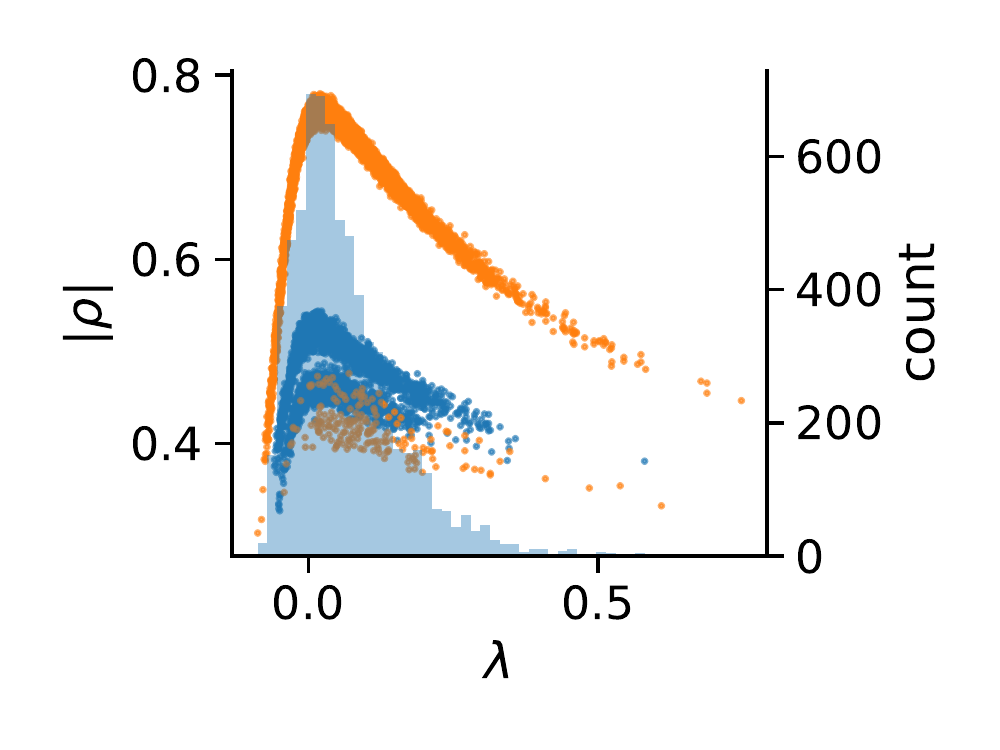}\\
    \end{minipage}
    \begin{minipage}{.36\textwidth}
    \centering (a)
    \end{minipage}
    \begin{minipage}{.31\textwidth}
    \centering (b)
    \end{minipage}
    \begin{minipage}{.31\textwidth}
    \centering (c)
    \end{minipage}
    \caption{     (a) Different decision functions
    used in the redundant control  variate removal. The weaker of any two control  variates is removed if
    the pair $(\bar\rho_{ij}, \rho_{ij})$ belongs to the shaded area of the considered function.
    The vertical dashed line indicates $\rho_{\min}$.
    (b,c) The absolute correlation of different constraints {to $V$}
    arising from different choices of $\lambda$. The blue dots represent constraints
    based on first order moments, while the orange refers to control  variates derived from second
    order moments. In both cases $10,\!000$ samples were used with
    30 initial samples for $\lambda$ from $N(0,1)$ and $k_{\min}=2$. A quadratic decision bound was used for
    the redundancy removal. Furthermore, a histogram of control  variates selected by Algorithm~\ref{alg:ssa}
    is given.
    In (b) $\E{X^A_2}$ in the dimerization model was estimated.
    In (c) $\E{X^X_{50}}$ in the processive modification model was estimated.
    }
    \label{fig:bdim}
\end{figure}

Checking both conditions is based on the correlation $\rho_{ij}$ between the $i$-th and $j$-th control variate
and the correlation $\rho_{iv}$ of a control variate $i$ to $V$.
The first condition is a simple lower threshold $\rho_{\min}$ for a correlation $\rho_{iv}$.
This condition aims to remove those variates  from the control  variate set that are only weakly correlated to $V$ (line~\ref{line:weak_covs}).
The rationale is that, if variate $i$ has a low correlation with the variable of interest $V$,
its computation may not be worth the costs.
Here, we propose to set $\rho_{\min}$ heuristically as
$$\rho_{\min} = \min\left(0.1, \frac{\max_{i}\rho_{iv}}{k_{\min}}\right)\,,$$
where $k_{\min}>1$ is an algorithm parameter.

The second condition aims to remove redundant conditions.
This is not only beneficial for the efficiency of the estimator, but also 
necessary for the matrix inversion \eqref{eq:lcv_asym}
because perfectly and highly correlated constraints will make
the covariance matrix estimate $\hat{\Sigma}_Z$ (quasi-) singular.
For all considered criteria we iterate over all tuples $(i,j)\in \{1,\dots,k\}^2$, $i\neq j$,
removing the weaker of the two, i.e.\ $\argmin_{k\in\{i,j\}}\rho_{kv}$,
if the two control  variates are considered redundant (line~\ref{line:redundant_covs}).

There are many ways to define such a redundancy criterion.
Here, we focus on criteria that are defined in terms of the
average correlation $\bar\rho_{ij}=(\rho_{iv} + \rho_{jv})/2$.
For two variates $i$ and $j$ we then check if their mutual correlation
$\rho_{ij}$ exceeds a some function $\phi$ of ${\bar{\rho}}_{ij}$,
i.e.\ we check the inequality
$$\phi(\bar\rho_{ij}) \leq \rho_{ij}\,.$$
If this inequality holds, constraint $\argmin_{k\in\{i,j\}}\rho_{kv}$ is removed.
Naturally, there are many possible choices for the above decision boundary $\phi$ (cf.\ Fig.~\ref{fig:bdim}a).

The simplest choice is to ignore $\bar\rho_{ij}$ and just fix a constant close to 1 as a threshold,
e.g.\ $\phi_c(\bar\rho_{ij})=.99$. While this often leads to the strongest variance reduction
and avoids numerical issues in the control variate computation, it turns out
that the computational overhead is not as well-compensated as by other choices of $\phi$ (see Section~\ref{sec:study}).

Another option is to fix a simple linear function, i.e.\ $\phi_{\ell}(\bar\rho_{ij})=\bar\rho_{ij}$.
For this choice the intuition is, that one of two constraints is removed if their mutual
correlation exceeds their average correlation with $V$.

Here, we also assess two quadratic choices for $\phi$. The first choice of $\phi_{q}(\bar\rho)=1 - (1-\bar\rho)^2$
is more tolerant than the linear function and more strict than a threshold function, except for
highly correlated control  variates.
Another variant of $\phi$ is given by including the lower bound $\rho_{\min}$ and scaling
the quadratic function accordingly: $\phi_{sq}(\bar\rho)=1-((1-\bar\rho)/(1-\rho_{\min}))^2$.
The different choices of $\phi$ considered here are plotted in Figure~\ref{fig:bdim}a.

\begin{algorithm}[t]
    \caption{\label{alg:ssa}Estimate the mean of species $i$ at time $T$}
    \begin{algorithmic}[1]
        \Procedure{EstimateMean}{$n, d, n_{\max}, n_{\lambda}, k_{\min}$}
        \State $L=\{\lambda_i\sim \pi_\lambda \mid 1\leq i< n_{\lambda} \} \cup \{ 0 \}$\label{line:lambda_sample}
        \State $P \leftarrow \{ ({m}, \lambda) | 1\leq\lvert {m} \rvert \leq n_{\max},
            \lambda\in L\}$\label{line:init_covs}
        \For{$i=1,\dots,n$}\label{line:main_loop}
            \State $\tau \leftarrow$ \textsc{SSA}($\pi_0,T,A$)\label{line:run_ssa}
            \State compute constraint values using $A$ and update $\hat\Sigma$ and $\hat{V}_i$\label{line:updates}
            \If{$i \mod d =0$}\label{line:cond_ck_begin}
                \State $\rho_{\min}\leftarrow \min\left(0.1,{\max_i \rho_{iv}}/{k_{\min}}\right)$
                \State $P\leftarrow P\setminus\{({m}_k, \lambda_k)\mid \rho_{kv} < \rho_{\min}\}$\label{line:weak_covs}
                \State $P\leftarrow P\setminus\{({m}_k, \lambda_k)\mid \exists i,j.i\neq j,
                \phi(\bar\rho_{ij}) < \rho_{ij},
                k=\argmin_{k\in\{i,j\}}\rho_{kv}\}$\label{line:redundant_covs}
            \EndIf
            \State remove unneeded accumulators from $A$\label{line:cond_ck_end}
        \EndFor
        \State \textbf{return} $\hat{V}_n - {(\hat{\Sigma}_{{Z}}^{-1}\hat\Sigma_{{Z}V})}^{\top}\hat{{Z}}_n$\label{line:compute_lcv}
        \EndProcedure
    \end{algorithmic}
\end{algorithm}

Now, we discuss  the choice of the initial control  variates. We identify control  variate $k$ by
a tuple $({m}_k, \lambda_k)$ of a moment vector ${m}_k$
and a time-weighting
parameter $\lambda_k$.
That is, we use $w(t)=e^{\lambda_k(T-t)}$ and $f(x)=x^{m_k}$ in \eqref{eq:poly_con}.
For a given set of parameters $L$, we use all moments up to some fixed order $n_{\max}$ (line~\ref{line:init_covs}).
The ideal set of parameters $L$ is generally not known. 
For certain choices the correlation of the control  variates and the variable of interest is
higher then for others.
To illustrate this, consider the above example of the birth-death process.
Choosing $\lambda=-\delta$ leads to a control variate that has a correlation of $\pm 1$ with
$V$. Therefore, the ideal
choice of initial values for would be $L=\{-\delta\}$.
This, however, is generally not known.
Therefore, we sample a set of $\lambda$'s
from some fixed distribution $\pi_{\lambda}$ (line~\ref{line:lambda_sample}).


\section{Case Studies}\label{sec:study}
We first define a criterion of \emph{efficiency} in order to estimate
whether the reduction in variance is worth the increased cost.
A natural baseline of a variance reduction is, that it is more efficient to pay for
the overhead of the reduction than to generate more samples to achieve a similar
reduction of variance.
Let $\sigma_Y^2$ be the variance of $Y$.
The \emph{efficiency} of the method is the ratio of the necessary
cost to achieve a similar reduction with the CV estimate $Y_{\text{CV}}$ compared to
the standard estimate $Y$~\cite{l1994efficiency}, i.e.
\begin{equation}\label{eq:efficiency}
E=\frac{c_0\sigma_Y^2}{c_1\sigma^2_{Y_{\text{CV}}}}\,.
\end{equation}
That ratio $c_0/c_1$ depends on both the specific implementation and the technical setup.
The cost increase is  mainly due to the computation of the integrals in \eqref{eq:simpl_exp}.
But the repeated checking of control  variates for efficiency also increases the cost.
The accumulation over the trajectory directly increases the cost of a single simulation
which is the critical part of the estimation.
To estimate the base-line cost $c_0$, 2000 estimations were performed without
considering any control variates.

The simulation is implemented in the Rust programming language\footnote{\url{https://www.rust-lang.org}}.
The model description is parsed from a high level specification. 
Rate functions are compiled to stack programs for fast evaluation.
Code is made available online \cite{cme-simulation-github}.

We consider four non-trivial case studies. Three models exhibit complex multi-modal behaviour.
We now describe the models and the estimated quantities in detail.

The first model is a simple dimerization on a countably infinite state-space.
\begin{model}[Dimerization]
We first examine a simple dimerization model on an unbounded state-space
$$\varnothing\xrightarrow{10}M,\quad 2M\xrightarrow{0.1}D$$
with initial condition $X_0^M=0$.
\end{model}
Despite the models simplicity, the moment equations are not closed for this system
due to the second reaction which is non-linear.
Therefore a direct analysis of the expected value would require a closure.
For this model we will estimate $\E{X^M_2}$.

The following two models are bimodal, i.e.\ they each posses two stable regimes
among which they can switch stochastically.
For both models we choose the initial conditions such that the process
will move towards either attracting region with equal probability.
\begin{model}[Distributive Modification]\label{model:dm}
The distributive modification model was introduced in \cite{cardelli2012cell}.
It consists of the reactions
\begin{align*}
    X + Y \xrightarrow{.001} B + Y\,,\quad
    &B + Y \xrightarrow{.001} 2 Y\,,\\
   Y + X \xrightarrow{.001} B + X\,,\quad
   &B + X \xrightarrow{.001} 2 X
\end{align*}
with initial conditions $X^X_0=X^Y_0=X^B_0=100$.
\end{model}



\begin{model}[Exclusive Switch]\label{model:es}
The exclusive switch model consists of 5 species, 3 of which are typically binary (activity states of the genes) \cite{loinger2007stochastic}.
\begin{center}
\begin{tabular}{c@{\hskip 1.5em}c@{\hskip 1.5em}c@{\hskip 1.5em}c@{\hskip 1.5em}c}
    $P_1 \xrightarrow{} \varnothing$&
    $G \xrightarrow{} G + P_2$&
    $G + P_1 \xrightarrow{} G_1$&
    $G_1 \xrightarrow{} G + P_1$&
    $G_1 \xrightarrow{} G_1 + P_1$\\
    \rule{0pt}{4ex} $P_2 \xrightarrow{} \varnothing$&
    $G \xrightarrow{} G + P_1$&
    $G + P_2 \xrightarrow{} G_2$&
    $G_2 \xrightarrow{} G + P_2$&
    $G_2 \xrightarrow{} G_2 + P_2$
\end{tabular}
\end{center}
with initial conditions $X^G_0=1$ and $X^{G_1}_0=X^{G_2}_0=X^{P_1}_0=X^{P_2}_0=0$.
\end{model}

We evaluate the influence of algorithm parameters, choices of distributions
to sample $\lambda$ from, and the influence of the sample size on the efficiency of
the proposed method.
Note that the implementation does not simplify the constraint representations
or the state space according to stoichiometric invariants or limited
state spaces.
Model~\ref{model:dm}, for example has the invariant $X^X_t+X^Y_t+X^B_t=\mathrm{const.}$, $\forall t\geq 0$, which
could be used to reduce the state-space dimensionality to two.
In Model~\ref{model:es} the invariant $\forall t\geq 0.X^G_t,X^{G_1}_t,X^{G_2}_t\in\{0,1\}$ could
be used to optimize the algorithm by eliminating redundant moments, e.g.\ $\expSym({(X^G)}^2)=\E{X^G}$.
Such an optimization would further increase the efficiency of the algorithm.

\begin{figure}[t]
    \centering
    \begin{minipage}{.3\textwidth}
    \centering
    \includegraphics[scale=.4]{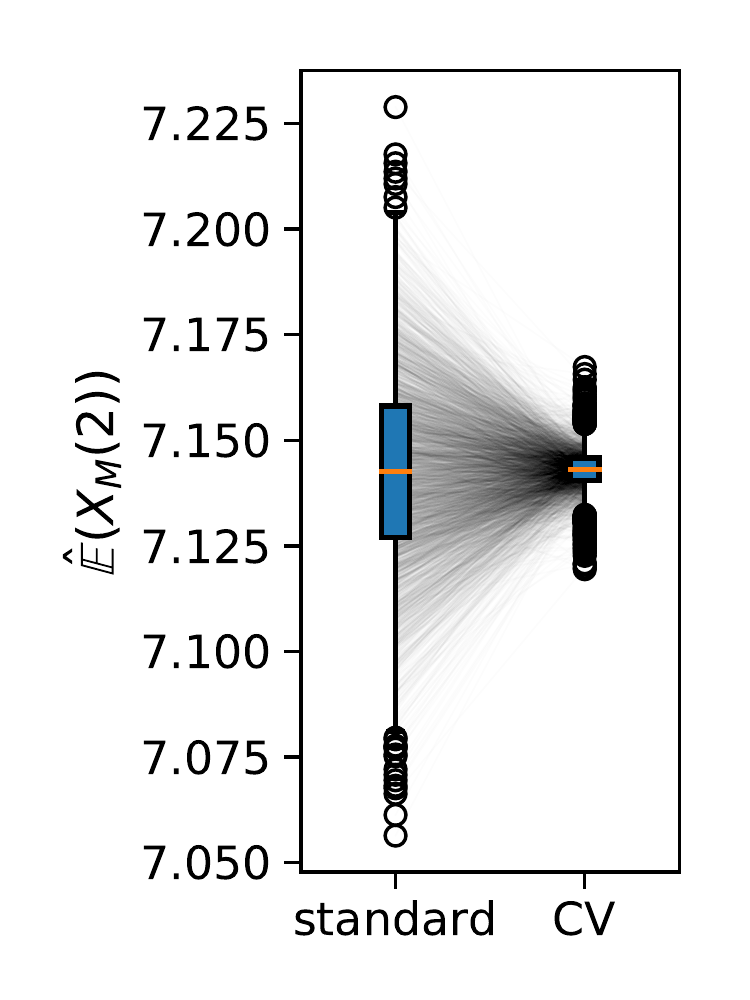}
    \end{minipage}
    \hspace{1em}
    \begin{minipage}{.5\textwidth}
    \centering
    \includegraphics[scale=.4]{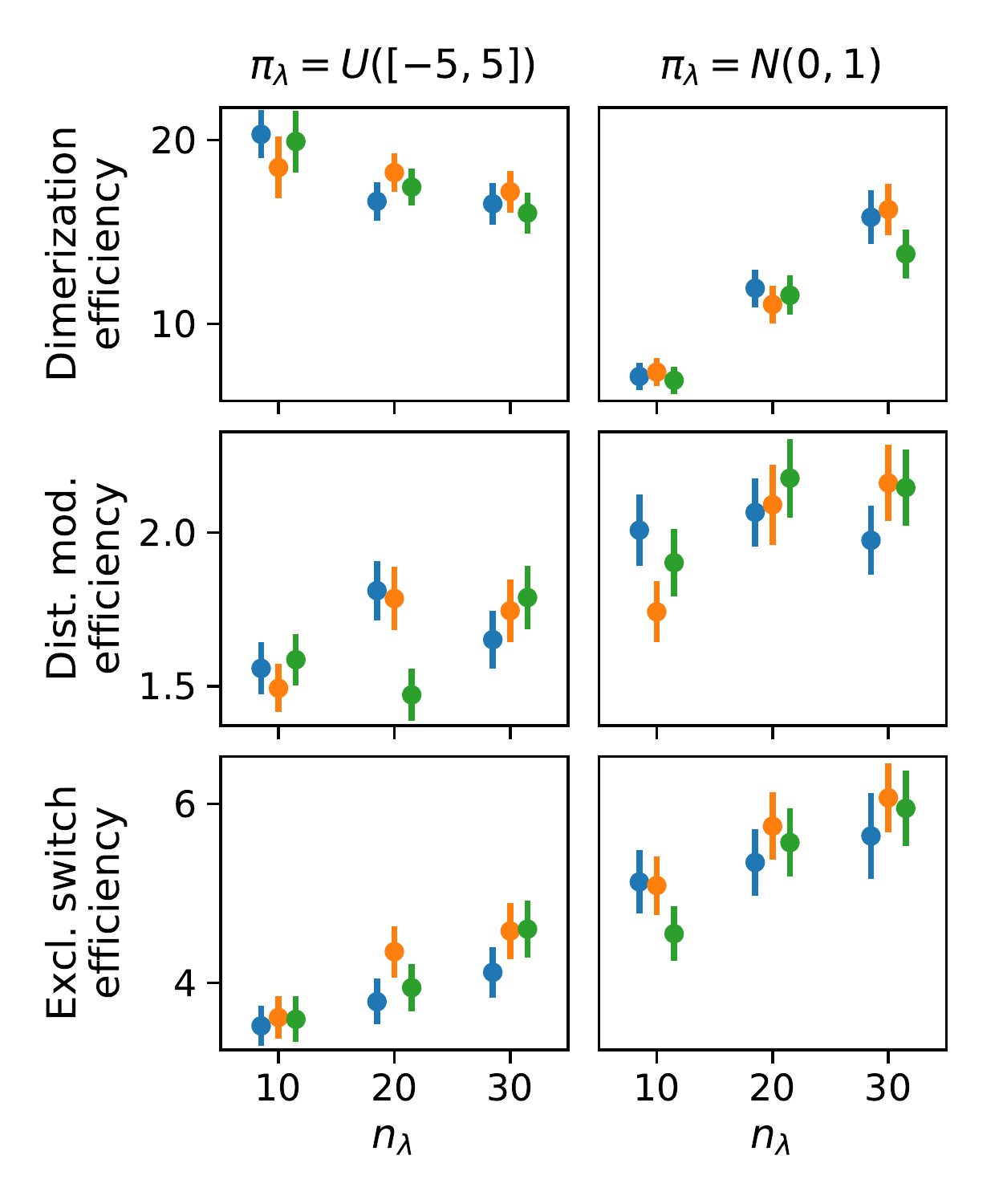}
    \end{minipage}
    \hspace{-2em}
    \begin{minipage}{0.12\textwidth}
    \includegraphics[scale=.5]{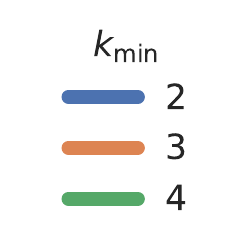}
    \end{minipage}
    \begin{minipage}{.3\textwidth}
    \hspace{2.8cm}(a)
    \end{minipage}
    \begin{minipage}{.69\textwidth}
    \centering(b)
    \end{minipage}
    \caption{(a) The effect of including control variates (CV) on the mean estimates
    $\hat{\mathbb{E}}(X^M_2)$ in the dimerization case study. Parameters
    were ${\pi}_{\lambda}=N(0,1)$, ${n}_{\lambda}=30$, ${k}_{\min}=4$, $\phi(\bar\rho)=1-{(1 - \bar\rho)}^2$.
    (b) The efficiencies for different valuations of ${n}_{\lambda}$ and ${k}_{\min} $
    and choices of ${\pi}_{\lambda}$. The sample size was $n=10,\!000$ in all cases
    with $d=100$.
    The bars give the 
    bootstrapped ($1000$ iterations)
    standard deviations.}
    \label{fig:efficiencies_prior}
\end{figure}

We first turn to the choice of the $\lambda$ sampling distribution. Here we
consider two choices:
\begin{enumerate}
    \item a standard normal distribution $N(0,1)$,
    \item a uniform distribution on $[-5,5]$.
\end{enumerate}
We deterministically include $\lambda=0$ in the constraint set, as this parameter
corresponds to a uniform weighting function.
We performed estimations on the case studies using different valuations of the
algorithm parameters of the minimum threshold $k_{\min}$ and
the number of $\lambda$-samples $n_{\lambda}$.
We used samples size $n=10,\!000$ and checked the control  variates every $d=100$ iterations
for the defined criteria.
For each valuation 1000 estimations were performed.
In Figure~\ref{fig:efficiencies_prior}b, we summarize the efficiencies
for the arising parameter combinations on the three case studies.
Most strikingly, we can note that the efficiency was consistently larger than one in
all cases.
Generally, the normal sampling distribution out-performed the 
alternative uniform distribution, except in case of the dimerization.
The reason for this becomes apparent, when examining Figure~\ref{fig:bdim}b,c:
In case of the dimerization model the most efficient constraints are found for
$\lambda\approx -3$, while in case of the distributive modification they are located
 just above 0 (we observe a similar pattern for the exclusive switch case study).
Therefore the sampling of efficient $\lambda$ values is more likely
using a uniform distribution for the dimerization case study, than it is for the others.
Given that larger absolute values for $\lambda$ seem unreasonable due 
their exponential influence on the weighting function and the problem of fixing
a suitable interval for a uniform sampling scheme, the choice of a standard normal
distribution for $\pi_{\lambda}$ seems superior.
\begin{figure}[t]
    \centering    
    \begin{minipage}{.9\textwidth}
    \centering
    \includegraphics[scale=.4]{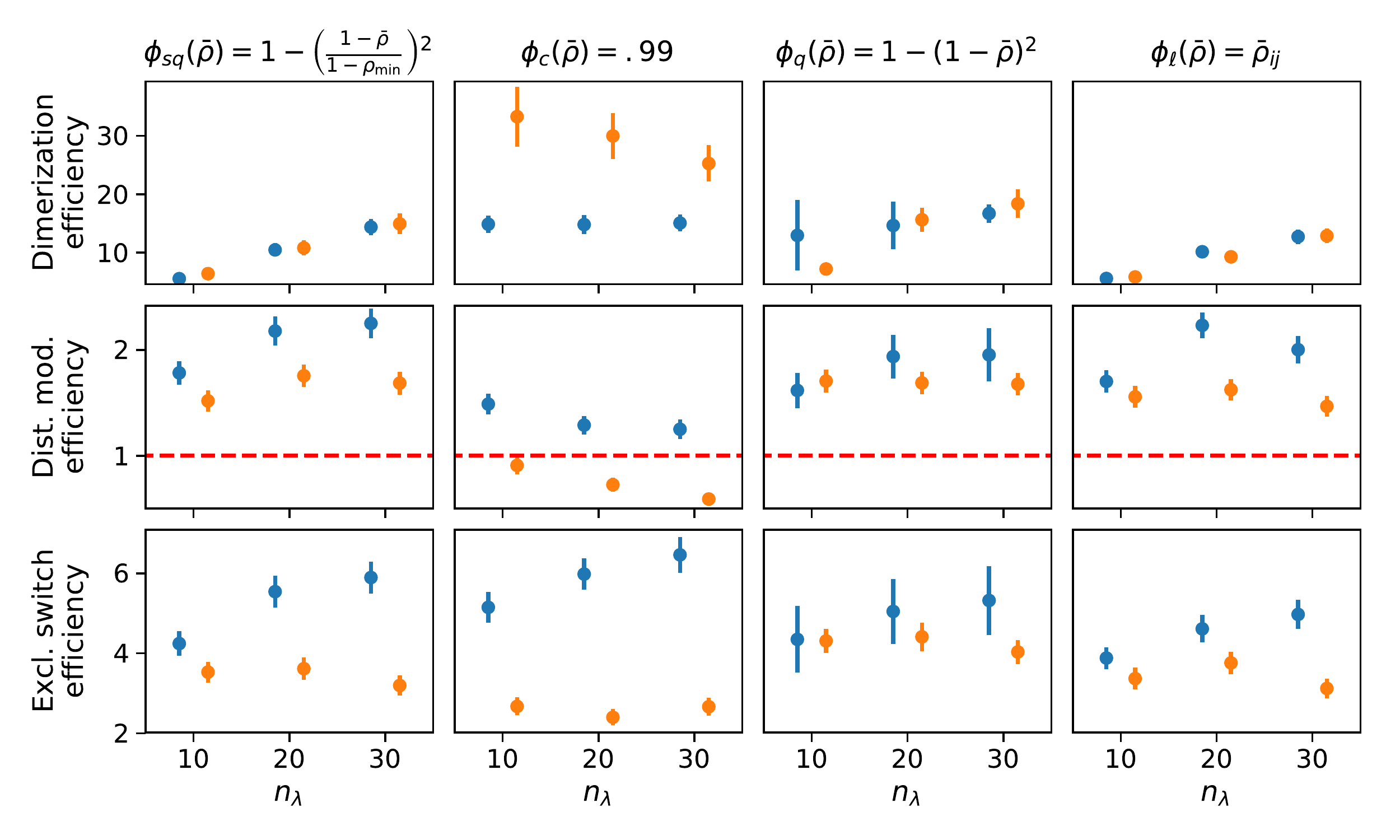}
    \end{minipage}\hspace{-1em}
    \begin{minipage}{0.09\textwidth}
    \centering
    \includegraphics[scale=.55]{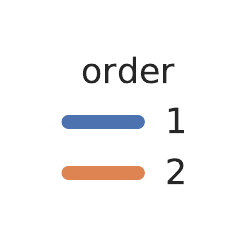}
    \end{minipage}
    \caption{The efficiency for different redundancy policies $\phi$ and maximal 
    moment orders $n_{\max}$. The sample size was $n=10,\!000$ in all cases
    with $d=100$. Furthermore, $k_{\min}=3$, $\pi_\lambda=N(0,1)$, and $n_{\max}=1$.
    The bars give the 
    bootstrapped (1000 iterations) standard deviations.\label{fig:efficiencies_order}}
\end{figure}

In Figure~\ref{fig:efficiencies_order} we compare efficiencies for
different maximum orders of constraints $n_{\max}$.
This comparison is performed for different choices
of the redundancy rule and initial $\lambda$ sample sizes $n_{\lambda}$.
Again, for each parameter valuation 1000 estimations were performed.
With respect to the maximum constraints order $n_{\max}$ we see a clear
tendency, that the inclusion of second order
constraints lessens the efficiency of the method.
In case of a constant redundancy threshold it even dips below break-even
for the distributive modification case study.
This is not surprising, since the inclusion of second order moments
increases the number of initial constraints quadratically
and the incurred cost, especially of the first iterations,
lessens efficiency.

\begin{figure}[tb]
    \centering
    \hfill
    \begin{minipage}{.72\textwidth}
    \includegraphics[scale=.4]{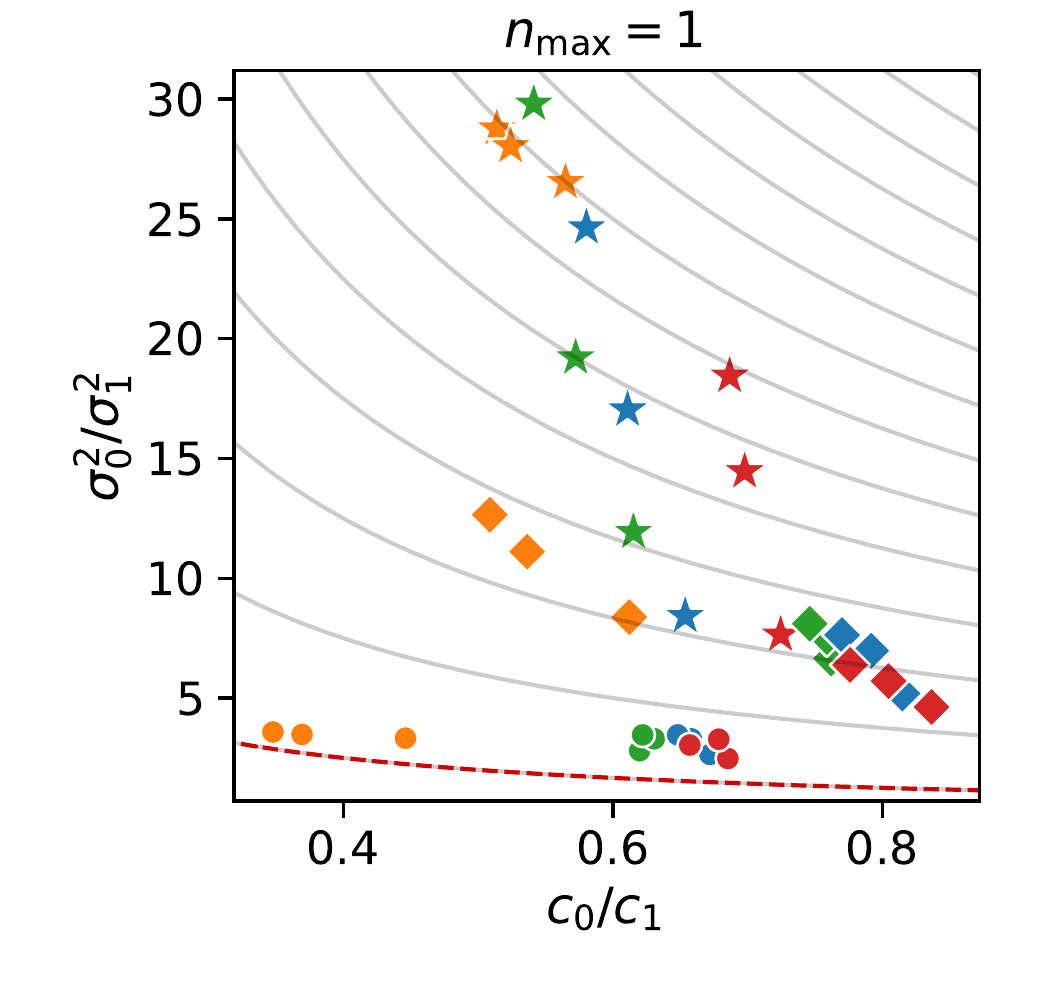}
    \includegraphics[scale=.4]{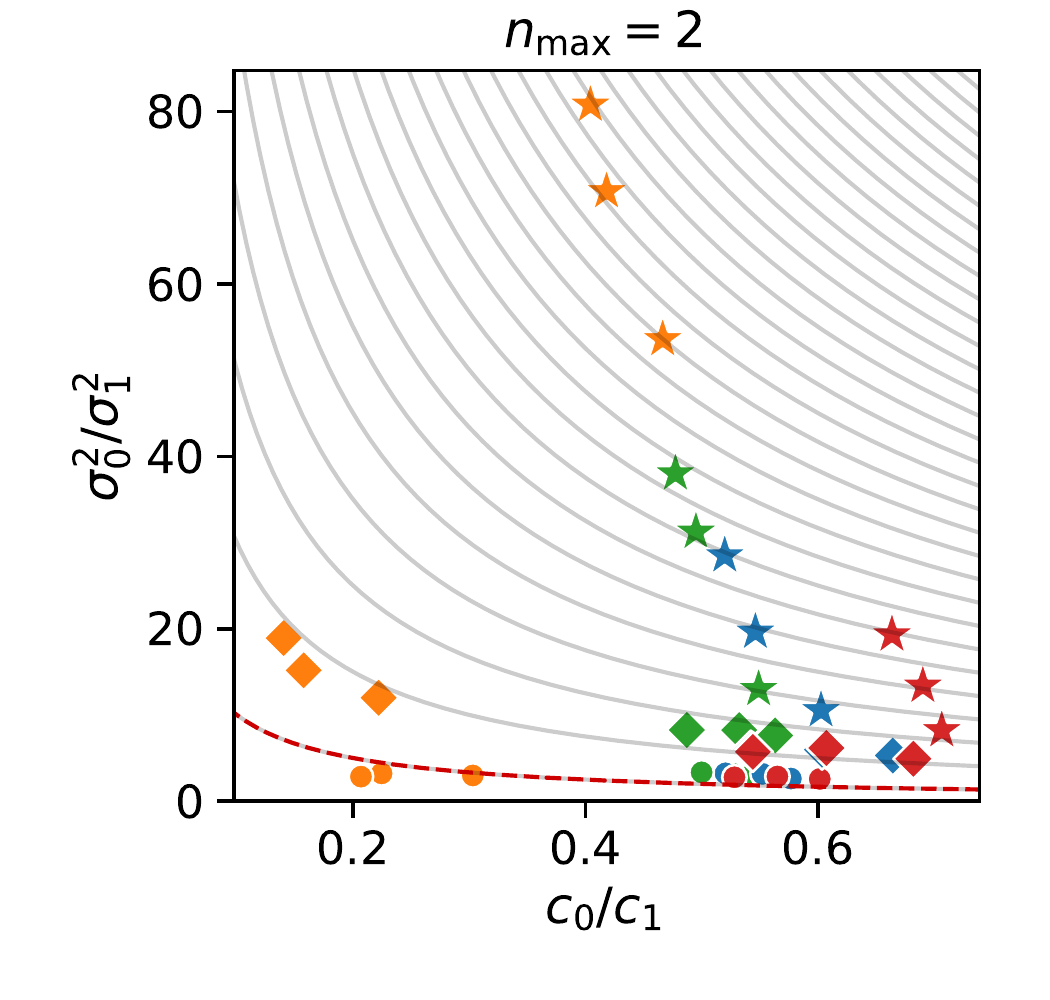}
    \end{minipage}
    \begin{minipage}{.18\textwidth}
    \includegraphics[scale=.4]{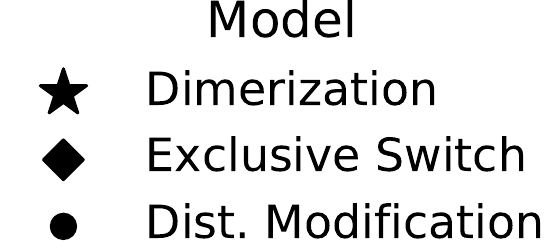}\\
    \includegraphics[scale=.4]{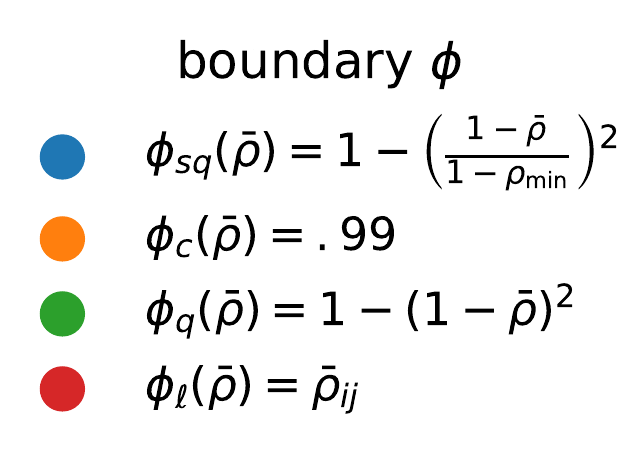}
    \vfill
    \end{minipage}
    \hfill
    \caption{A visualisation of the trade-off between variance reduction $\sigma_0^2/\sigma_1^2$
    and cost ratio $c_0/c_1$. Isolines for efficiencies are given in grey. The break-even
    is marked by the dashed red line.
    Markers of the same kind differ in $n_{\lambda}$ and shift with increasing value
    upwards in variance reduction and lower in $c_0/c_1$, i.e.\ the shift is to the left and upwards
    with increasing $n_{\lambda}$.
    The sample size was $n=10,\!000$ in all cases
    with $d=100$. Furthermore, $k_{\min}=3$ and $\pi_\lambda=N(0,1)$.}
    \label{fig:trade_off}
\end{figure}

Figure~\ref{fig:trade_off} depicts the trade-off between the variance reduction $\sigma_0^2/\sigma_1^2$
versus the cost ratio $c_0/c_1$. Comparing the redundancy criterion based on a constant threshold
$\phi_c$ to the others,
we observe both a larger variance reduction and an increased cost. This is due to the fact, that
more control  variates are included throughout the simulations (cf.\ Appendix~\ref{sec:table_app},
Tables~\ref{tab:eff1},\ref{tab:eff2}). Depending on the sample distribution $\pi_{\lambda}$
and the case study, this permissive strategy may pay off. In case of the dimerization, for example,
it pays off, while in case of the distributive modification it leads to a lower efficiency ratio.
In the latter case the model is more complex, and therefore the set of initial
control  variates is larger. With a more permissive redundancy strategy, more control  variates are kept
(ca.\ 10 when using $\phi_c$ vs.\ ca.\ 2--3 for the others).
The other redundancy boundaries move the results further in the direction of less variance reduction
while keeping the cost increase low.
On the opposite end is the linear $\phi_{\ell}$.
The quadratic versions $\phi_{q}$ and $\phi_{sq}$ can be found in the middle of this spectrum.

We also observe, that an increase of $n_{\lambda}$ is particularly beneficial, if
the sampling distribution $\pi_{\lambda}$ does not capture the parameter region of
the highest correlations well.
This can be seen for the Dimerization case study, where the variance reduction increases
strongly with increasing sample size (App.~\ref{sec:table_app} Tables~\ref{tab:eff1},\ref{tab:eff2}).
Since $\pi_{\lambda}=N(0,1)$, more samples are needed to sample efficient $\lambda$-values (cf.\ Figure~\ref{fig:bdim}b).

\begin{figure}[t]
    \centering
    \hfill
    \begin{minipage}{.48\textwidth}
    \centering
    \includegraphics[scale=.4]{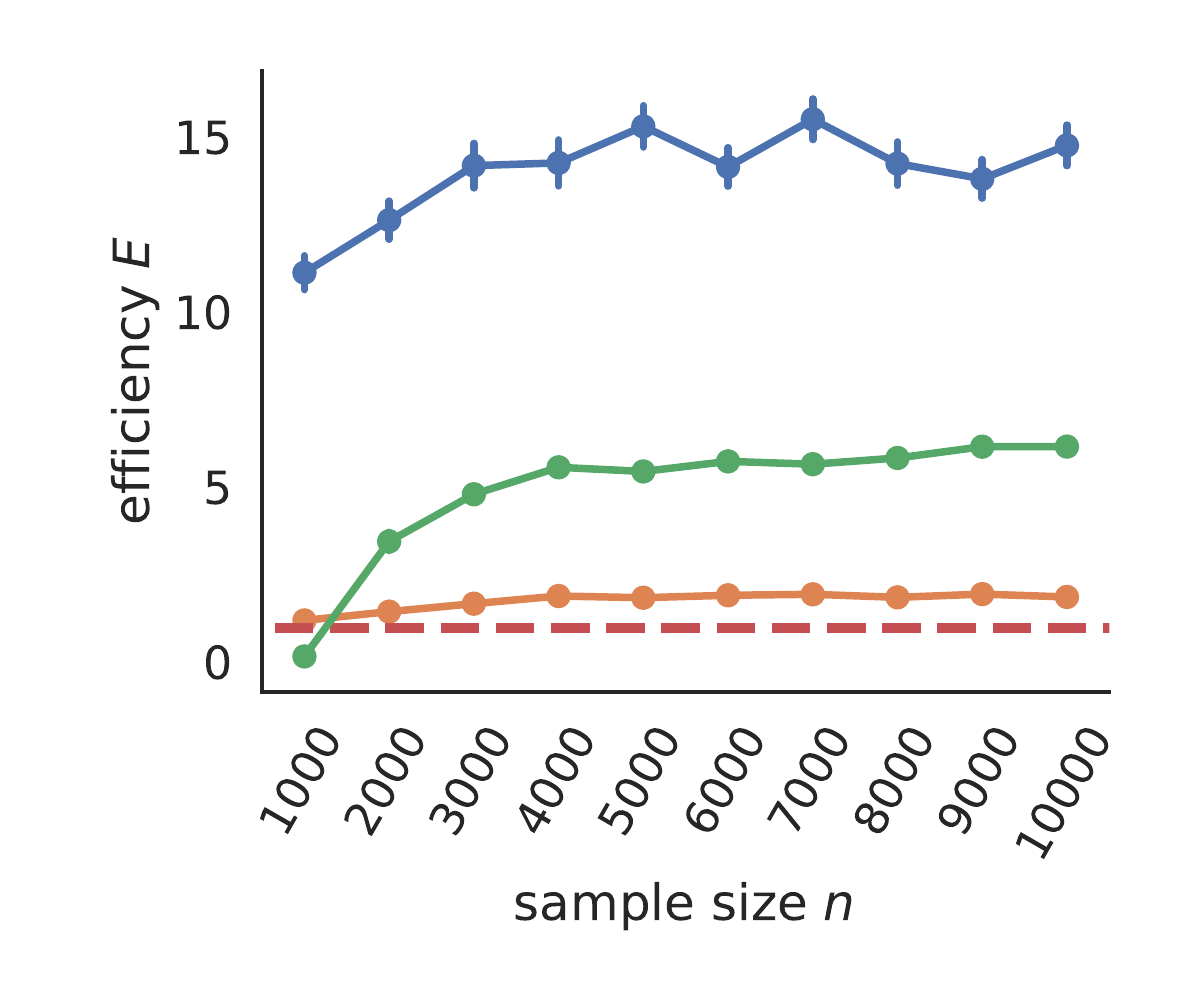}
    \end{minipage}
    \hfill
    \begin{minipage}{.51\textwidth}
    \includegraphics[scale=.4]{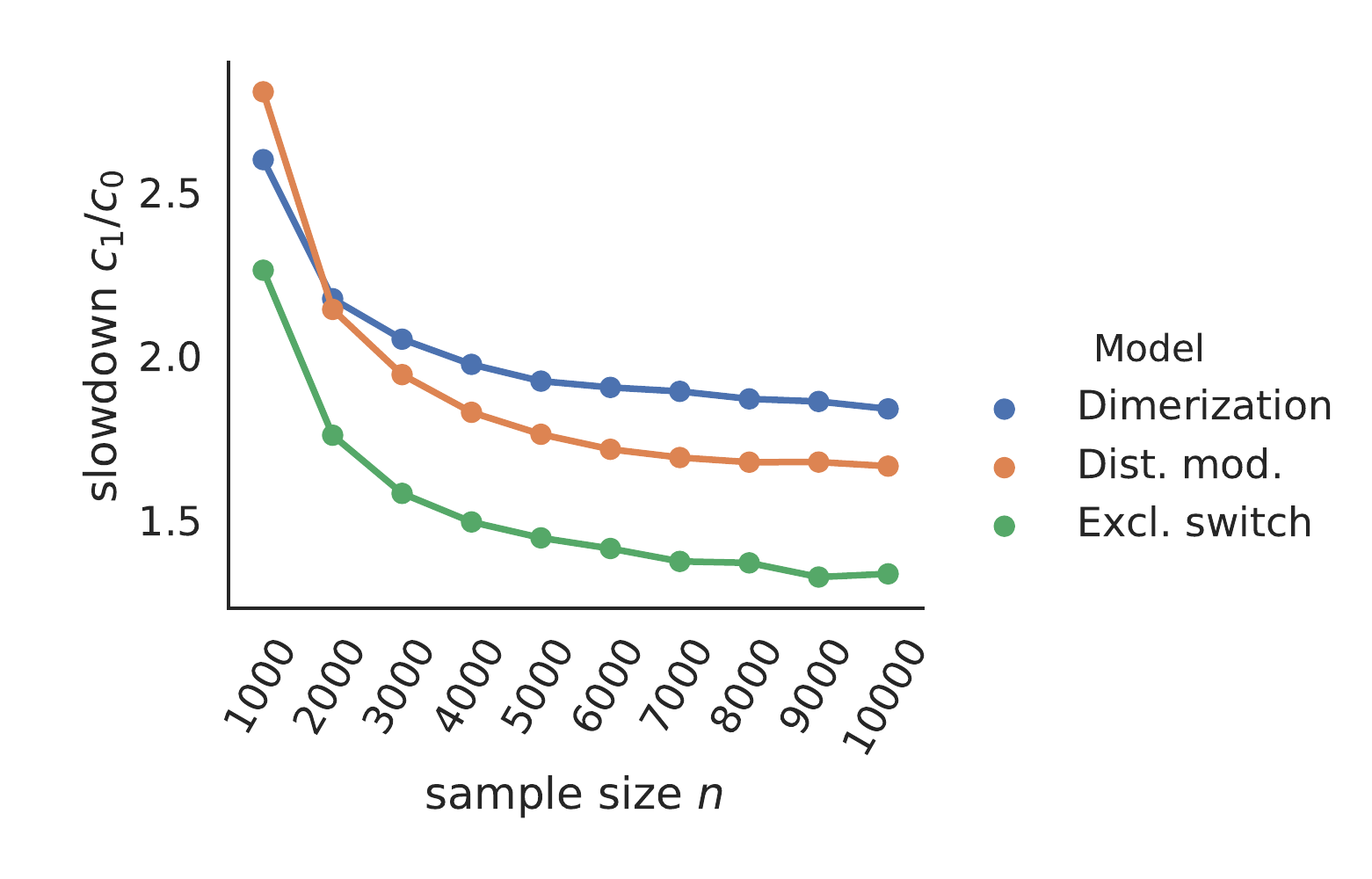}
    \end{minipage}
    \hfill
    \begin{minipage}{.48\textwidth}
    \centering
    \hspace{1.5em}(a)
    \end{minipage}
    \begin{minipage}{.51\textwidth}
    \centering
    \hspace{2em}(b)
    \end{minipage}
    \caption{
    (a) The effect of sample size on the efficiency $E$ in the different case studies.
    The break-even $E=1$ is marked by the dashed red line.
    (b) The cost increase due to the variance reduction over different sample sizes.}
    \label{fig:sample_sizes}
\end{figure}

Finally, we discuss the effect of the sample size $n$ on the efficiency $E$. In Figure~\ref{fig:sample_sizes}
we give both the efficiencies and the slowdown for different sample sizes.
As a redundancy rule we used the unscaled quadratic function, 30 initial values of $\lambda$,
and $k_{\min}=3$. With increasing sample size, the efficiency usually approaches an upper limit.
This is due to the fact that most control  variates are dropped early on and
the control  variates often remain the same for the rest of the simulations.
If we assume there are no helpful control  variates in the initial set
and all would be removed at iteration 100, the efficiency would approach 1
with $n\to \infty$.

\section{Conclusion}\label{sec:conclusion}
In this work we have shown that known constraints on the moment dynamics can be successfully
leveraged in simulation-based estimation of expected values.
The empirical results indicate that
the supplementing a standard SSA estimation with moment information
can drastically reduce the estimators' variance.
This reduction is paid for by accumulating information on the trajectory
during simulation.
However, the reduction is able to compensate for this increase.
This means that for fixed costs, using   estimates with
  control variates is more beneficial than using estimates without control variates.

While a variety of algorithmic variants was evaluated, many aspects remain subject to
further study.
In particular different choices of $f$ and $w$ in \eqref{eq:poly_con} may improve 
efficiency further.
These choices become particularly interesting when moving from the estimation
of simple first order moments to more complex queries such as behavioural probabilities
of trajectories.
In such cases, one might even attempt to find efficient control variate functions
using machine learning methods.

Another open question regarding this work is its performance when multiple
quantities instead of a single quantity are to be estimated. In such
a case, constraints would be particularly beneficial, if they lead
to improvements  as many estimation targets as possible.

Furthermore the identification of the best weighting parameters $\lambda$
could be done in a more adaptive fashion.
The presented scheme of a sampling from $\pi_{\lambda}$ could be extended
into a Bayesian-like procedure, wherein the values for $\lambda$ are periodically
re-sampled from a distribution that is adjusted according to the best-performing
constraints up to that point.

\subsubsection{Acknowledgements} This work was supported by the DFG project MULTIMODE.

%
%

\bibliographystyle{splncs04}
\bibliography{paper.bib}

\newpage
\appendix
\section{Supplementary Material}\label{sec:table_app}
\begin{figure}[h]
    \centering    
    \begin{minipage}{.9\textwidth}
    \centering
    \includegraphics[scale=.4]{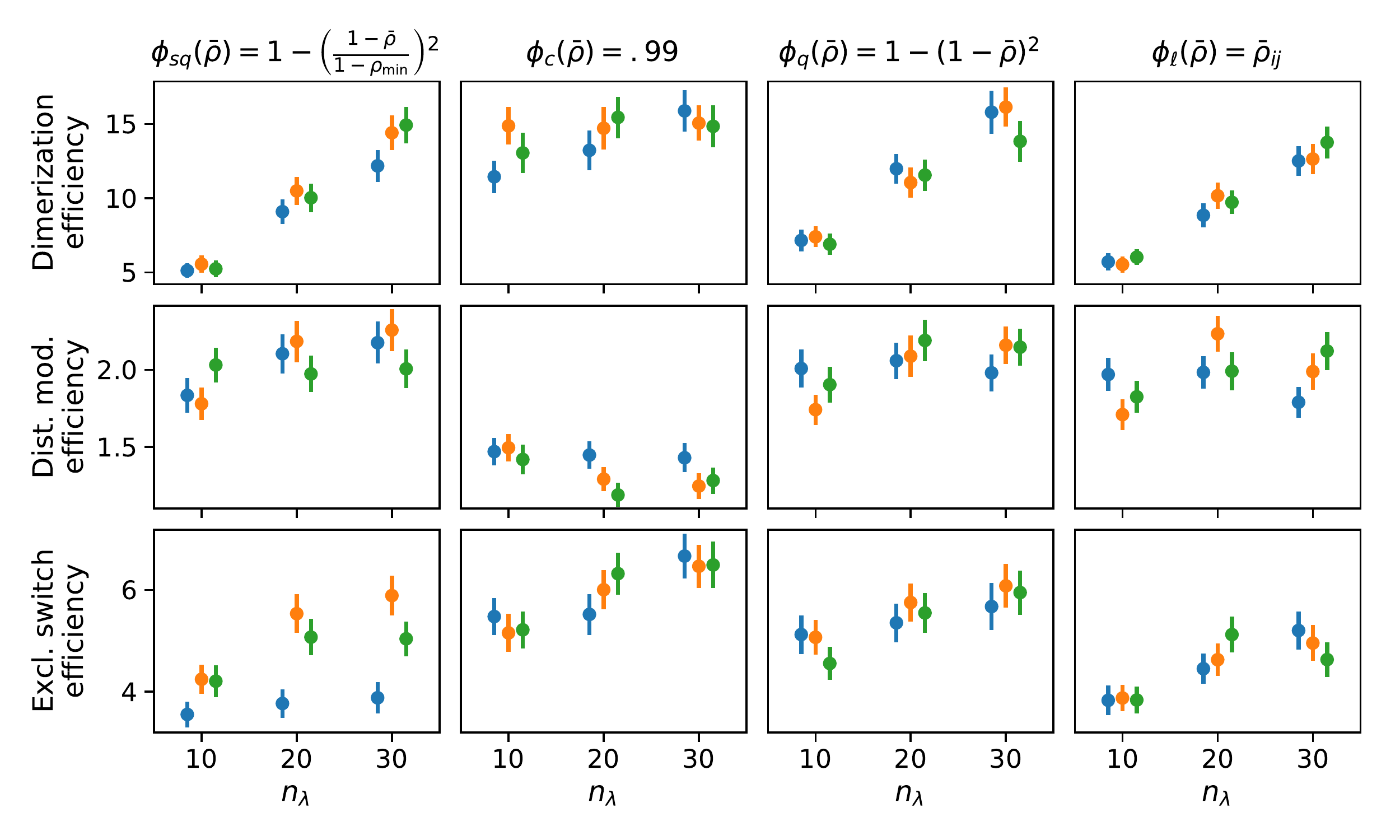}
    \end{minipage}
    \begin{minipage}{0.09\textwidth}
    \includegraphics[scale=.55]{legend.pdf}
    \end{minipage}
    \caption{The empirical efficiencies for different $n_\lambda$ and $k_{\min} $.
    On the considered case studies. The sample size was $n=10,\!000$ in all cases
    with $d=100$.
    1000 estimations were performed for each case.
    The bars give the 
    bootstrapped (1000 iterations) standard deviations.
    The break-even $E=1$ is marked by the dotted red line.\label{fig:efficiencies_alg_params}}
\end{figure}

In Figure~\ref{fig:efficiencies_alg_params} we give detailed information on the influence of
algorithm parameters $k_{\min}$, the number of initial $\lambda$ values, and
different redundancy rules.
The $\lambda$ sampling distribution $\pi_{\lambda}$ is a standard normal.
\begin{table}[]
    \centering
\begin{tabular}{l@{\hskip 12pt}l@{\hskip 12pt}l@{\hskip 12pt}r@{\hskip 12pt}r@{\hskip 12pt}r@{\hskip 12pt}r}
\toprule
  Case study &   $n_{\lambda}$ & $\phi$ & $1-\frac{\sigma_1^2}{\sigma_0^2}$ &  slowdown &  efficiency& $\lvert P\rvert$ \\\midrule
Dimerization & 10 &$\phi_{sq}$ &  0.881641 &  1.530137 &    5.558692 &   1.621917 \\
            &    &$\phi_{c}$ &  0.965224 &  1.945588 &   14.859417 &   3.338501 \\
            &    &$\phi_{q}$ &  0.916445 &  1.625232 &    7.409904 &   1.997045 \\
            &    & $\phi_{\ell}$ &  0.868288 &  1.380344 &    5.529745 &   1.081152 \\\cmidrule{2-7}
            & 20 &$\phi_{sq}$ &  0.941153 &  1.637272 &   10.437978 &   1.842971 \\
            &    &$\phi_{c}$ &  0.964204 &  1.907999 &   14.747328 &   2.915082 \\
            &    &$\phi_{q}$ &  0.947984 &  1.747519 &   11.072422 &   2.227250 \\
            &    & $\phi_{\ell}$ &  0.931030 &  1.433401 &   10.169570 &   1.088572 \\\cmidrule{2-7}
            & 30 &$\phi_{sq}$ &  0.959517 &  1.723449 &   14.404936 &   1.972426 \\
            &    &$\phi_{c}$ &  0.962514 &  1.770936 &   15.142156 &   2.216103 \\
            &    &$\phi_{q}$ &  0.966216 &  1.847441 &   16.117387 &   2.446661 \\
            &    & $\phi_{\ell}$ &  0.945724 &  1.456432 &   12.710196 &   1.084188 \\\midrule
Dist.\ mod. & 10 &$\phi_{sq}$ &  0.619560 &  1.488483 &    1.770218 &   3.232575 \\
            &    &$\phi_{c}$ &  0.700255 &  2.241695 &    1.492171 &   8.008607 \\
            &    &$\phi_{q}$ &  0.643550 &  1.613001 &    1.743500 &   3.817641 \\
            &    & $\phi_{\ell}$ &  0.596650 &  1.459405 &    1.703170 &   2.657000 \\\cmidrule{2-7}
            & 20 &$\phi_{sq}$ &  0.697414 &  1.519425 &    2.181687 &   2.631677 \\
            &    &$\phi_{c}$ &  0.713445 &  2.706546 &    1.292838 &  10.295856 \\
            &    &$\phi_{q}$ &  0.697654 &  1.585313 &    2.092817 &   3.398235 \\
            &    & $\phi_{\ell}$ &  0.695846 &  1.473976 &    2.235418 &   2.226530 \\\cmidrule{2-7}
            & 30 &$\phi_{sq}$ &  0.712941 &  1.543068 &    2.263644 &   2.378037 \\
            &    &$\phi_{c}$ &  0.721354 &  2.874249 &    1.252541 &  10.910880 \\
            &    &$\phi_{q}$ &  0.711877 &  1.607712 &    2.164485 &   2.979704 \\
            &    & $\phi_{\ell}$ &  0.669963 &  1.522184 &    1.996300 &   2.085473 \\\midrule
Excl. switch & 10 &$\phi_{sq}$ &  0.807184 &  1.227471 &    4.239255 &   2.536479 \\
            &    &$\phi_{c}$ &  0.880285 &  1.633530 &    5.135205 &   7.411732 \\
            &    &$\phi_{q}$ &  0.849082 &  1.312416 &    5.067770 &   3.639250 \\
            &    & $\phi_{\ell}$ &  0.783459 &  1.195821 &    3.874778 &   2.090101 \\\cmidrule{2-7}
            & 20 &$\phi_{sq}$ &  0.856593 &  1.263340 &    5.539683 &   2.206154 \\
            &    &$\phi_{c}$ &  0.910480 &  1.864405 &    6.011256 &   9.441336 \\
            &    &$\phi_{q}$ &  0.867987 &  1.317958 &    5.765884 &   3.140806 \\
            &    & $\phi_{\ell}$ &  0.825518 &  1.243075 &    4.627662 &   1.981143 \\\cmidrule{2-7}
            & 30 &$\phi_{sq}$ &  0.869165 &  1.298893 &    5.905196 &   2.059415 \\
            &    &$\phi_{c}$ &  0.921019 &  1.966191 &    6.461331 &   9.928998 \\
            &    &$\phi_{q}$ &  0.876822 &  1.340409 &    6.079876 &   2.762449 \\
            &    & $\phi_{\ell}$ &  0.843288 &  1.288925 &    4.968796 &   1.983174 \\
\bottomrule
\end{tabular}
    \caption{$n_{\max}=1$, $n=10,\!000$, $d=100$, $k_{\min}=3$}
    \label{tab:eff1}
\end{table}

\begin{table}[]
    \centering
\begin{tabular}{l@{\hskip 12pt}l@{\hskip 12pt}l@{\hskip 12pt}r@{\hskip 12pt}r@{\hskip 12pt}r@{\hskip 12pt}r}
\toprule
  Case study &   $n_{\lambda}$ & $\phi$ & $1-\frac{\sigma_1^2}{\sigma_0^2}$ &  slowdown &  efficiency &$\lvert P\rvert$ \\\midrule
Dimerization & 10 &$\phi_{sq}$ &  0.905254 &  1.659799 &    6.402491 &   2.388472 \\
            &    &$\phi_{c}$ &  0.987526 &  2.474939 &   33.074955 &   6.501180 \\
            &    &$\phi_{q}$ &  0.923063 &  1.822654 &    7.195544 &   3.179257 \\
            &    & $\phi_{\ell}$ &  0.878232 &  1.415909 &    5.830248 &   1.092264 \\\cmidrule{2-7}
            & 20 &$\phi_{sq}$ &  0.949038 &  1.831995 &   10.797164 &   2.890898 \\
            &    &$\phi_{c}$ &  0.985710 &  2.391457 &   29.704344 &   5.450299 \\
            &    &$\phi_{q}$ &  0.968076 &  2.021487 &   15.662368 &   3.681229 \\
            &    & $\phi_{\ell}$ &  0.925413 &  1.449386 &    9.298961 &   1.072761 \\\cmidrule{2-7}
            & 30 &$\phi_{sq}$ &  0.964855 &  1.924268 &   14.911787 &   3.026275 \\
            &    &$\phi_{c}$ &  0.981507 &  2.144089 &   25.520987 &   4.179125 \\
            &    &$\phi_{q}$ &  0.973902 &  2.095985 &   18.507746 &   3.685851 \\
            &    & $\phi_{\ell}$ &  0.948349 &  1.507425 &   12.904707 &   1.074538 \\\midrule
Dist.\ mod. & 10 &$\phi_{sq}$ &  0.619450 &  1.734737 &    1.519168 &   3.148184 \\
            &    &$\phi_{c}$ &  0.665361 &  3.301159 &    0.909443 &  13.456259 \\
            &    &$\phi_{q}$ &  0.680592 &  1.840457 &    1.705876 &   3.864240 \\
            &    &$\phi_{\ell}$ &  0.612674 &  1.662962 &    1.556868 &   2.659592 \\\cmidrule{2-7}
            & 20 &$\phi_{sq}$ &  0.684789 &  1.811408 &    1.755652 &   2.687379 \\
            &    &$\phi_{c}$ &  0.689835 &  4.455005 &    0.726640 &  17.609554 \\
            &    &$\phi_{q}$ &  0.687665 &  1.901258 &    1.688449 &   3.413595 \\
            &    &$\phi_{\ell}$ &  0.651262 &  1.770238 &    1.623924 &   2.266729 \\\cmidrule{2-7}
            & 30 &$\phi_{sq}$ &  0.690602 &  1.922217 &    1.686011 &   2.375455 \\
            &    &$\phi_{c}$ &  0.649191 &  4.837419 &    0.591701 &  19.145054 \\
            &    &$\phi_{q}$ &  0.701253 &  2.001179 &    1.677062 &   3.007525 \\
            &    &$\phi_{\ell}$ &  0.639123 &  1.894074 &    1.467403 &   2.086275 \\\midrule
Excl. switch & 10 &$\phi_{sq}$ &  0.811956 &  1.505521 &    3.544783 &   2.323999 \\
            &    &$\phi_{c}$ &  0.916866 &  4.507566 &    2.681363 &  21.692390 \\
            &    &$\phi_{q}$ &  0.868874 &  1.776190 &    4.309354 &   4.739893 \\
            &    & $\phi_{\ell}$ &  0.795802 &  1.466579 &    3.353046 &   2.016196 \\\cmidrule{2-7}
            & 20 &$\phi_{sq}$ &  0.832562 &  1.657484 &    3.617313 &   2.085711 \\
            &    &$\phi_{c}$ &  0.934280 &  6.348223 &    2.406431 &  29.976320 \\
            &    &$\phi_{q}$ &  0.878944 &  1.879341 &    4.416281 &   3.990881 \\
            &    & $\phi_{\ell}$ &  0.837922 &  1.647329 &    3.759896 &   1.978017 \\\cmidrule{2-7}
            & 30 &$\phi_{sq}$ &  0.829427 &  1.844766 &    3.190308 &   2.043201 \\
            &    &$\phi_{c}$ &  0.947324 &  7.130628 &    2.673225 &  32.513670 \\
            &    &$\phi_{q}$ &  0.878830 &  2.053317 &    4.034987 &   3.611746 \\
            &    & $\phi_{\ell}$ &  0.824936 &  1.838879 &    3.118728 &   1.978836 \\
\bottomrule
\end{tabular}
    \caption{$n_{\max}=2$, $n=10,\!000$, $d=100$, $k_{\min}=3$}
    \label{tab:eff2}
\end{table}
\end{document}